\documentclass[3p,times,pdflatex]{elsarticle}
\usepackage[breaklinks,hidelinks,colorlinks]{hyperref}

\usepackage{mathtools}
\usepackage{enumitem}
\usepackage{natbib}
\usepackage[dvipsnames]{xcolor}
\makeatletter
\renewcommand{\fnum@figure}{Fig. \thefigure}
\makeatother

\begin{document}
\begin{frontmatter}
%
%
%
\title{Performance of a triple-GEM detector with capacitive-sharing 3-coordinate (X--Y--U)-strip anode readout.}
%
%
\author[1]{Kondo Gnanvo\texorpdfstring{\corref{cor1}}{}} 
\texorpdfstring{\ead{kagnanvo@jlab.org}}{}
\cortext[cor1]{Corresponding author}
\author[1]{Seung Joon Lee}
\author[2]{Bertrand Mehl}
\author[2]{Rui de Oliveira}
\author[1]{Andrew Weisenberger}
\address[1]{Thomas Jefferson National Accelerator Facility, Newport News VA 23606, USA}
\address[2]{CERN Esplanade des Particules 1 P.O. Box 1211 Geneva 23}
\begin{abstract}
The concept of capacitive-sharing readout, described in detail in a previous study,  offers the possibility for the development of high-performance three-coordinates (X--Y--U)-strip readout for Micro Pattern Gaseous Detectors (MPGDs) using simple standard PCB fabrication techniques. Capacitive-sharing (X--Y--U)-strip readout allows  simultaneous measurement of the Cartesian coordinates x and y of the position of the particles together with a third coordinate u along the diagonal axis in a single readout PCB. This provides a powerful tool to address  multiple-hit ambiguity and enable pattern recognition capabilities in moderate particle flux environment of collider or fixed target experiments in high energy physics HEP) and nuclear physics (NP). We present in this paper the  performance of a 10 cm × 10 cm triple-GEM detector with capacitive-sharing (X--Y--U)-strip anode readout. Spatial resolutions of the order of $\sigmaup_{x}^{res}$ = 71.6 $\pm$ 0.8 $\muup$m  for X-strips, $\sigmaup_{y}^{res}$ = 56.2 $\pm$ 0.9 $\muup$m for Y-strips and  $\sigmaup_{u}^{res}$ = 75.2 $\pm$ 0.9 $\muup$m  for U-strips have been obtained at a beam test at Thomas Jefferson National Accelerator Facility (Jefferson Lab). Modifications of the readout design  of future prototypes to improve the spatial resolution and challenges in scaling to large-area MPGDs are discussed.

\end{abstract}
\begin{keyword}
GEM \sep capacitive-sharing readout \sep spatial resolution \sep strip readout \sep (X--Y--U)-strip 
%
\end{keyword}
\end{frontmatter}
%
%
\section{Introduction}
\label{sec:introduction}
Micro Pattern Gaseous Detector (MPGD) technologies such as Gas Electron Multiplier (GEM) \cite{SAULI:1997nim},  Micro-Mesh Gaseous Structures (Micromegas)  \cite{GIOMATARIS:199629}, or resistive micro-well ($\muup$RWELL) \cite{Bencivenni:2015} are  widely used or being considered for tracking, particle identification and calorimetry applications in current and future HEP and NP experiments. These detectors combine an electron multiplication device for the amplification of the electronic signal with anode readout printed circuit boards (PCBs) made of finely segmented strip or pad pick-up electrodes to allow precise measurement of particle position coordinates. As an example,  triple-GEM detectors used in several particle physics experiments \cite{COMPASS:2002, GNANVO:2015nim, GNANVO:2016nim} typically have two-dimensional (X--Y)-strip readouts with a strip pitch of 400 $\muup$m that provides spatial resolutions better than 60 $\muup$m for both x and y coordinates. For large-scale particle physics experiments where large-area trackers are needed, the requirement for fine segmentation strip readout to achieve ample spatial resolution will lead to a prohibitively large number of electronic channels to be read out. This has serious implications in terms of electronic and data acquisition costs, large data rate management as well as integration of the MPGD trackers inside the already complex detector system of large scale collider or fixed target experiments in NP and HEP. Integration challenges include cooling, shielding of front end electronics, cables routing, connectors choices that optimize detector acceptance, material thickness and radiation damages mitigation plans in high radiation environment. Each of these challenges increases in complexity with the number of electronic channels to be read out. \\In an attempt to reduce the level of complexity  while maintaining spatial resolution performance of MPGD detectors, several R\&D efforts~\cite{DIXIT2004721,ATTIE2020163286,paulColasResSh,azmoun:2020tns,azmoun:2022tns,Zhang:2017dqw} were initiated over the past ten years to develop low-channel counts, high-performance readout structures for large-area MPGDs. A novel approach based on a large pitch ($\ge$ 800 $\muup$m) capacitive-sharing readout structure was recently introduced  as the anode readout structure of a $\muup$RWELL detector~\cite{capaSh_urwell2022}. Spatial resolution performance of the order of 50 to 75 $\muup$m, comparable to that of a standard COMPASS (X--Y)-strip readout \cite{COMPASS:2002} with  400 $\muup$m pitch was demonstrated via test beam experiments. Capacitive-sharing structures are a cost-effective and high-performance anode readout option, well suited for GEM, Micromegas or $\muup$RWELL technologies and compatible with standard PCB fabrication techniques for large-area tracking detectors. \\ 
For large MPGD tracking detectors in particle physics experiments, even for low to moderate particle background rate environment, events overlapping in time will cause multiple hits in both X and Y strips of the detector readout layers, resulting in the so-called "ghost hits" or multiple-hit ambiguity when attempting to match hits coordinates in x and y for the 2D reconstruction of the particle positions. The probability of "ghost hits" increases  with the strip length and width and thus consequently with the large pitch readout structures under development~\cite{azmoun:2022tns,Zhang:2017dqw,capaSh_urwell2022}. One  promising approach to mitigate the impact of "ghost hits" is the development of a 3-coordinate readout structure such as the (X--Y--U)-strip readout presented in this work. In this configuration, the U-strips which are oriented at a 45$^{\circ}$ angle with respect to the X-strip axis, provide the third coordinate u, correlated in space and time to both x and y coordinates. The combination of capacitive-sharing  structures with the 3-coordinate readout offers the possibility for high-performance, low-channel count and cost-effective anode readout PCBs for large-area MPGDs.\\
We discuss in the current paper, the performance based on beam studies of a 10 cm $\times$ 10 cm triple-GEM prototype instrumented with a (X--Y--U)-strip capacitive-sharing readout. In section~\ref{sec:xyurwellproto}, we provide a detailed description of the readout design followed by a presentation in section~\ref{sec:beamtest} of the test beam setup at Jefferson Lab used for the performance study of the prototype. In sections~\ref{sec:characterization} and~\ref{sec:resolution}, the characteristic performances  and  spatial resolution performance of the prototype  are discussed. Finally, in section~\ref{sec:conclusion}, we introduce future R\&D ideas to improve  performance of 3-coordinates capacitive-sharing readout for tracking in future large experiments.
%

\section{Triple-GEM prototype  with (X--Y--U)-strip capacitive-sharing  readout} 
\label{sec:xyurwellproto}
A small (10 cm $\times$ 10 cm) triple-GEM prototype equipped with a 3-coordinate capacitive-sharing (X--Y--U)-strip anode readout was assembled and underwent beam tests at Jefferson Lab. The readout PCB  is based on a stack of five layers of copper-clad 50 $\muup$m thick polyimide (Kapton) layers. The basic concept of the structure is  described in detail in a previous work~\cite{capaSh_urwell2022}.
%
%
\begin{figure}[!ht]
\centering
\includegraphics[width=1.0\columnwidth,trim={0pt 65mm 0pt 0mm},clip]{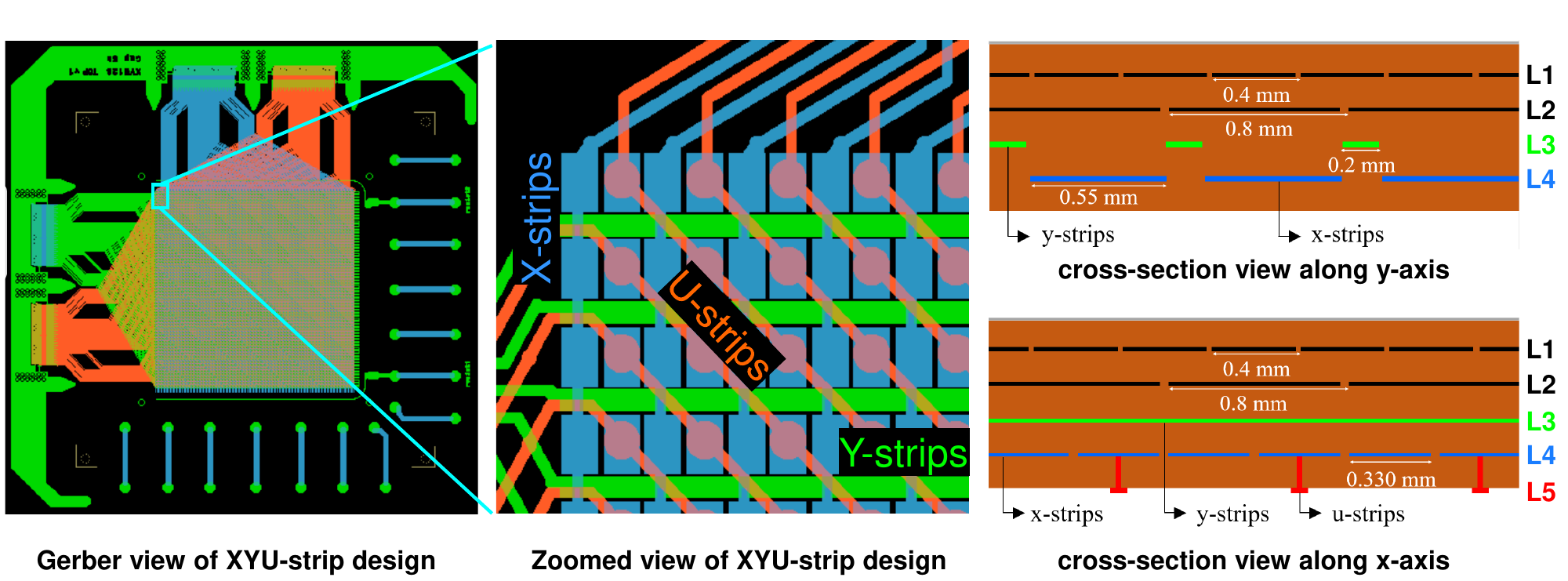}
\caption{\label{fig:capaShStripxyu} Left: Gerber file generated diagram of the capacitive-sharing PCB with a 3-coordinate (X--Y--U)-strip readout pattern. Center: zoomed in view of the strip structure, Right: cross section of the multi-layer PCB with the stack of two capacitive-sharing layers (black pads) and the two strip layers, Y-strips (green) in L3, X-strips (blue strips) and pads for U-strips (blue pads) in L4 and connecting the vias and traces (red) in layer L5.}
\end{figure}
%
%
The top side of each Kapton layer has a laminated 5 $\muup$m thick Cu-layer with a given pad or strip pattern. The Kapton foil serves as dielectric layer that facilitates the transfer of induced signals between the Cu-layers via capacitive coupling. The five layers of the readout PCB of the  prototype discussed in this work are color-coded in the pictures of Fig.~\ref{fig:capaShStripxyu}. The top two layers L1 and L2 are the capacitive layers, each having an array of pads (black) with a pitch of 400 $\muup$m and 800 $\muup$m respectively. The Y-strips layer (L3) has horizontal strips (green) for y-coordinates and layer L4 has both vertical strips (X-strips in blue) for x coordinates and a matrix of rectangular pads (blue) between X-strips that are connected together into U-strips along the diagonal axis through vias connecting to traces (shown in red) on layer L5 and routed to the readout connectors for the u coordinates. The naming of the Y-strips, X-strips and U-strips comes from the coordinate system in the test beam setup described  in section~\ref{sec:beamtest} for x,y, and u coordinates of the particle positions. Both Y-strips and X-strips have a pitch of 800 $\muup$m and a strip width of 200 $\muup$m and 330 $\muup$m respectively. The U-pads of L4 have a dimension of 330 $\muup$m $\times$  550 $\muup$m and a pitch along the diagonal axis equal to 800 $\muup$m $\times$ cos (45$^{\circ}$) =  567 $\muup$m. The traces in L5 have a width of 100 $\muup$m but could be made as narrow as 50 $\muup$m in future prototypes to minimize their contribution to the detector input capacitance and  cross-talk induced on X-strips.

\section{Study of the performance of XYU-GEM prototype in beam test at Jefferson Lab}
\label{sec:beamtest}
%
\subsection{Beam test setup in the Experimental Hall D at Jefferson Lab}
\begin{figure}[!ht]
\centering
\includegraphics[width=0.95\columnwidth,trim={0pt 10mm 0pt 5mm},clip]{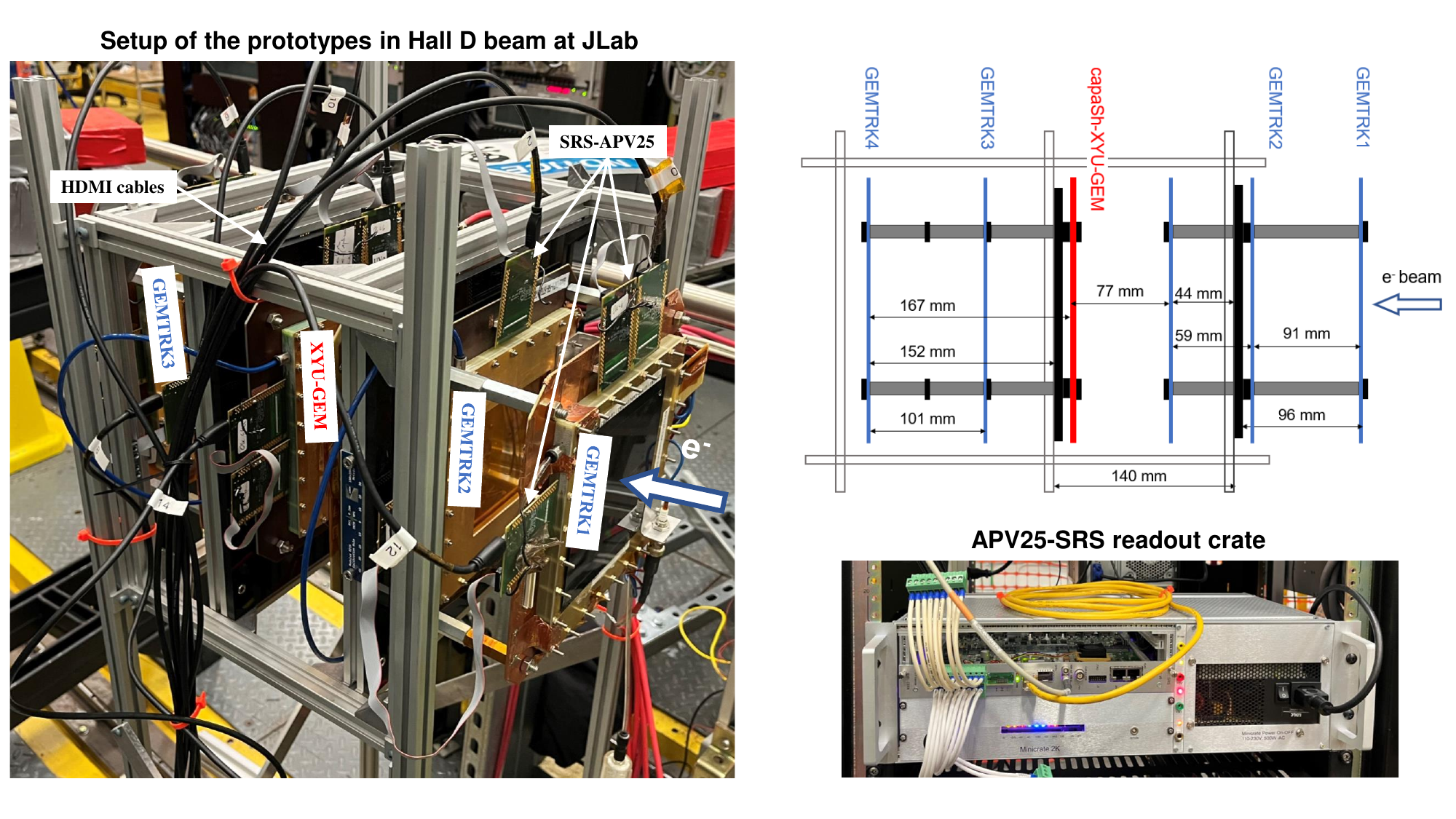}
\caption{\label{fig:beamtestsetup} Instrumentation setup to perform beam tests. Left: beam test telescope setup with four GEM trackers and the XYU-GEM prototype all instrumented with APV25-SRS FE cards, installed in the electron arm of the PS. Top-right: layout of the telescope. Bottom-right: photograph of the APV25-SRS readout crate.}
\end{figure}

The  prototype described in section~\ref{sec:xyurwellproto}, was installed in a tracking telescope setup in the electron arm of the Pair Spectrometer (PS) equipment of the GLUeX~\cite{gluex}  experiment in Hall D at Jefferson Lab and operated during the Fall 2022 run of the Continuous Electron Beam Accelerator Facility (CEBAF). The layout of the setup is shown on top right of Fig.~\ref{fig:beamtestsetup}. The tracking telescope is composed of two pairs of standard GEM trackers~\cite{COMPASS:2002}, one upstream and one downstream of the XYU-GEM prototype. The GEM trackers provide precision tracking for the study of spatial resolution and efficiency of the prototype under test. The PS delivers a clean electron beam from gamma ($\gamma$)  conversion of the GLUeX photon beam line at an energy ranging from 3 to 6 GeV and at a rate of $\sim$7 kHz, corresponding to the expected pair production rate for a CEBAF 300 nA beam current. The converted electrons are bent in the horizontal direction by a 1.8 T dipole to cover the full range in the horizontal plane of the detectors in the tracking telescope. The width of the beam in the vertical axis is ~2.45 mm. 
 All four GEM trackers, together with the XYU-GEM prototype were operated with a Ar:CO2 (70:30) gas mixture. A single HV channel of a CAEN N1470 power supply is coupled to a standard COMPASS resistive divider~\cite{COMPASS:2002} to distribute the voltages across the 7 electrodes of each triple-GEM detector. All detectors are instrumented with the APV25-based~\cite{apv25} Scalable Readout System (SRS)~\cite{srs2011} developed by CERN RD51 collaboration \cite{rd51coll}. The SRS-APV25 front end (FE) hybrids on the detectors are connected via HDMI cables to the SRS-ADC cards of the SRS crate \cite{srs2011}, shown in the bottom right of  Fig.~\ref{fig:beamtestsetup}. The SRS data acquisition (SRS-DAQ) has a sampling rate of 50 ns and nine APV25 time samples, corresponding to an acquisition window of 450 ns are recorded per readout channel in the current setup. The DAQ operates in a standalone mode independently from the GLUeX main DAQ system to keep the data under $\leq$1 kHz in order to cope with the limited performance of the DAQ computer used during the beam test. DATE and AMORE software packages, both developed by the ALICE collaboration at CERN~\cite{aliceSoft}, were used for the DAQ and online monitoring respectively.  The monitoring and analysis software tool used for the APV25-SRS electronics was developed in the AMORE framework and is called amoreSRS~\cite{amoresrs}. It includes scripts for decoding the APV25 raw data, applying common mode correction,  pedestal offset subtraction and zero suppression.
%
%
%
\subsection{SRS-APV25 gain calibration}
\label{subsec:gaincomparison}
\begin{figure}[!ht]
\centering
\includegraphics[width=0.9\columnwidth,trim={0pt 35mm 0pt 25mm},clip]{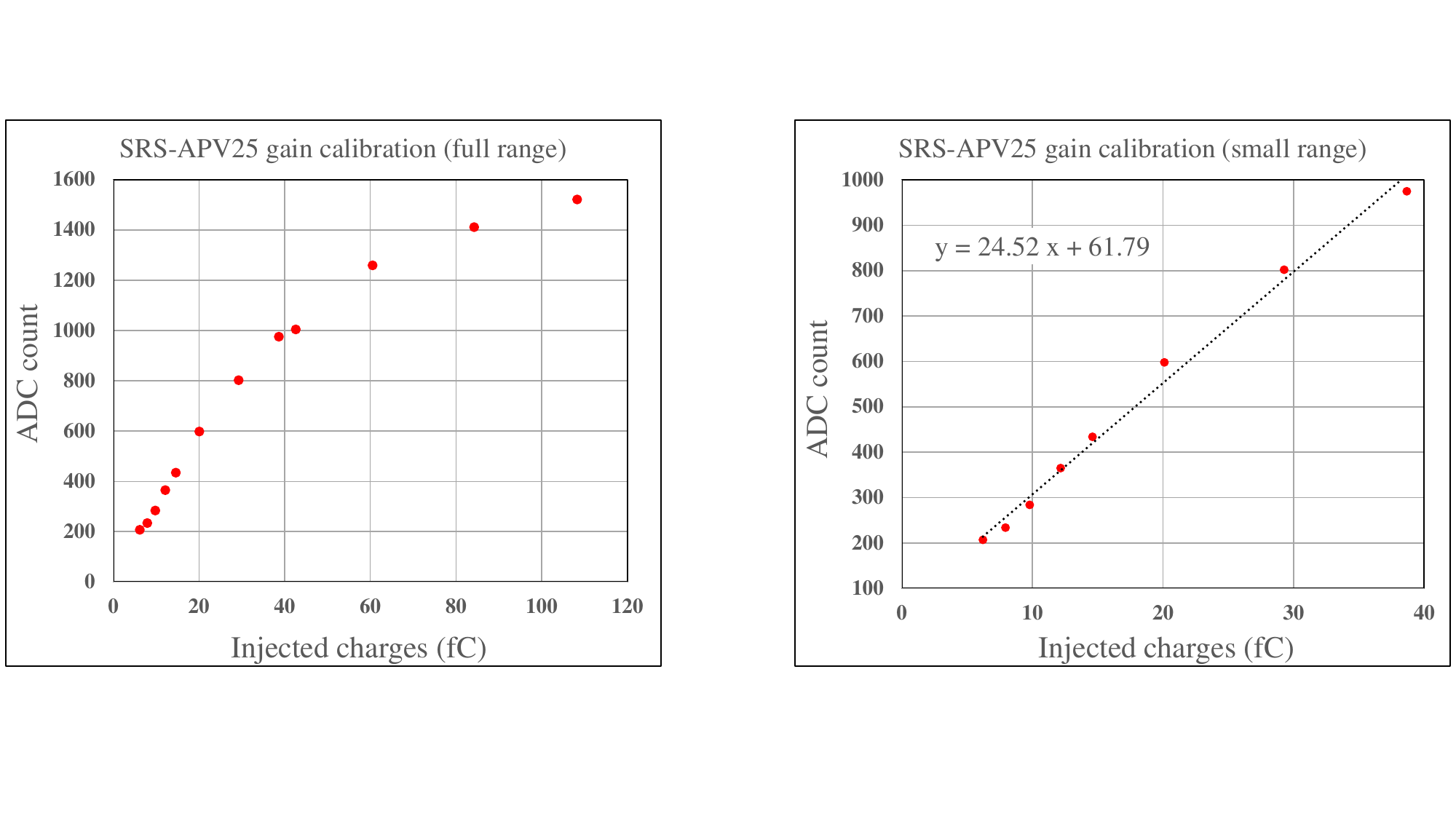}
\caption{\label{fig:apv_gain} SRS-APV25 gain calibration plots. Left: ADC count vs. injected charges (fC) in a single APV channel. \textit{Left}: full range of the gain curve showing response non linearity above $\sim$40 fC; \textit{Right}: the limited range of interest for the X--Y--U GEM signal shows good linearity up to 40 fC.}
\end{figure}

The gain calibration of the SRS-APV25 signal amplitude as a function of the injected charge has been studied and documented in~\cite{apvgain}. The experimental setup for the gain measurement for the default settings and configuration of the SRS implementation of the APV25 electronics are also described in detail in the document. The plots of Fig.~\ref{fig:apv_gain} show the ADC channel of the SRS-APV25 as a function of the injected charge on a single strip. The plot on the left is the APV25 gain curve over a wide range of injected signal, up to 100 fC and the response becomes non-linear starting at $\sim$40 fC corresponding to an injected charge of  $\sim$250 000 electrons. On the right, the plot demonstrates the limited range of the injected charge in which the APV25 response is linear up to to 1000 ADC channels. 
The effective gain of a typical triple-GEM detector operating with a Ar:CO$_2$ 70:30 gas mixture is $\sim$8 $\times \,10^3$ for an average voltage across the GEM foils equal to 376V~\cite{COMPASS:2002}, equal to a total charge of $\sim$280 000 electrons at the front end electronics. If we assume equal sharing between all 3 sets of strips, each set will collect  $\sim$94 000 electrons. In this case with a cluster size equal to 5 and assuming the central strips collect about 60\% of the total charge or $\sim$56 000 electrons thus a little less than 9 fC. This corresponds to 300 ADC channels and is thus within the linear range of the SRS-APV25 readout electronics. 
%
%
\subsection{Electron beam characteristics}
\begin{figure}[!ht]
\centering
\includegraphics[width=0.95\columnwidth,trim={0pt 0mm 0pt 0mm},clip]{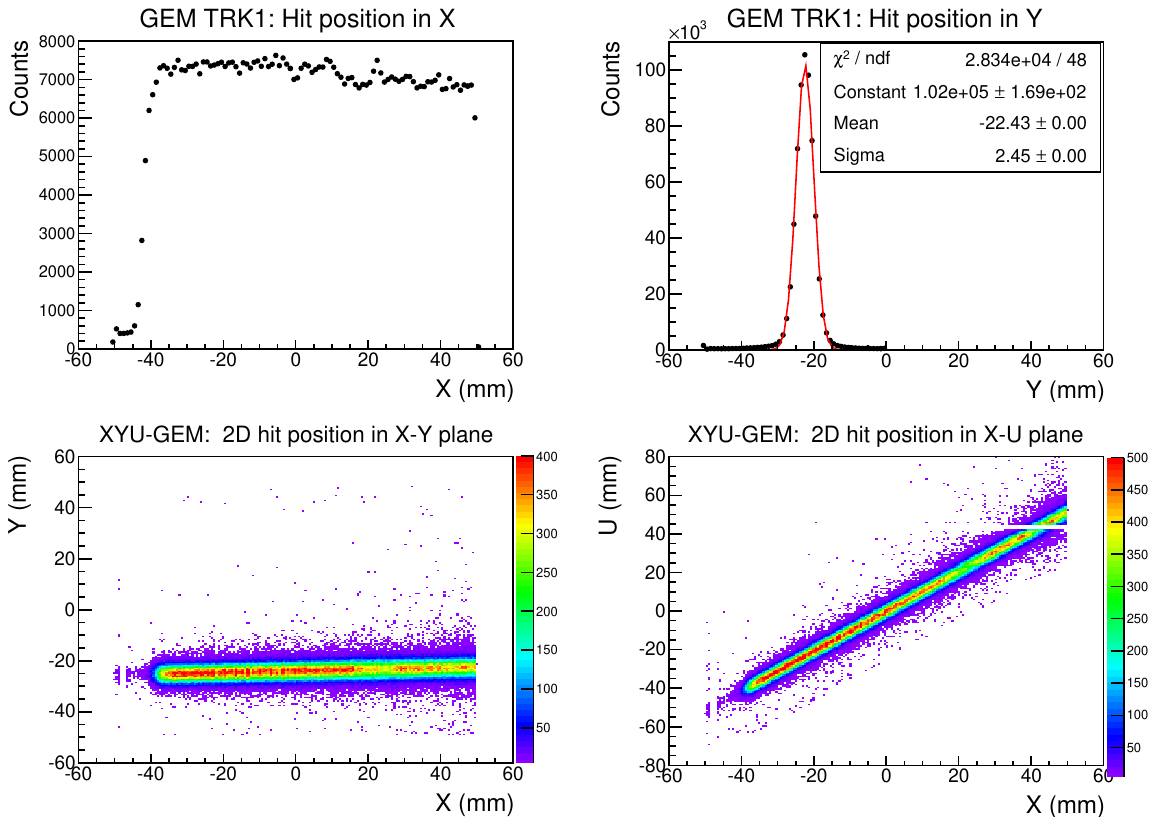}
\caption{\label{fig:beamTestProfile} Electron beam profile from Hall D Pair Spectrometer measured by the first tracker (GEMTRK1). Top: The 1D distribution of the hit coordinates in x \textit{(left)} and in y \textit{(right)}.  Bottom:  Reconstruction of the 2D profile of the PS electron beam in the X-Y plane \textit{(left)} and the X-U plane \textit{(right)} of the XYU-GEM prototype under study.}
\end{figure}

The characteristics of the electron beam during the beam tests were determined. For each triggered event, an electron track is reconstructed when hits are recorded simultaneously in both the x and y axis of each of the four GEM trackers. The plots in the top of Fig.~\ref{fig:beamTestProfile} display a uniform distribution of the reconstructed PS electron beam hit profile along the x-axis and narrow profile along y-axis fitted to a Gaussian with a width $\sigmaup$ of 2.54 mm.
The 2D profiles of the PS electron beam obtained from the reconstructed hit positions in the X-Y and X-U planes of the XYU-GEM prototype are shown at the left and right respectively at the bottom of Fig.~\ref{fig:beamTestProfile}. Note that the reconstructed u-coordinates from the U-strips along the diagonal axis are plotted as a function of the x-coordinates as seen in the right plot.
%
%
%
\section{Performance characteristics of the XYU-GEM prototype}
\label{sec:characterization}
High voltage scans were performed to study the performance of the (X--Y--U)-strip readout as a function of the detector gain. Absolute gain measurements and gain curves as a function of applied voltage for standard triple-GEM detectors are well documented in the literature \cite{COMPASS:2002,BACHMANN2001548} and are not the focus of the study of the XYU-GEM prototype  in this paper. The gain of the detector is inferred from the average voltage across the GEM foils of the XYU-GEM prototype based on the ADC gain calibration studies of section~\ref{subsec:gaincomparison}.  Performance characteristics such as the efficiency, the cluster charge (in ADC counts), the strip multiplicity and the charge sharing between strips as a function of the voltage are discussed in detail in the following subsections. 
%
%
\subsection{Strip pedestal noises }
\label{subsec:pedestals}
\begin{figure}[!ht]
\centering
\includegraphics[width=1. \columnwidth,trim={0pt 0mm 0pt 0mm},clip]{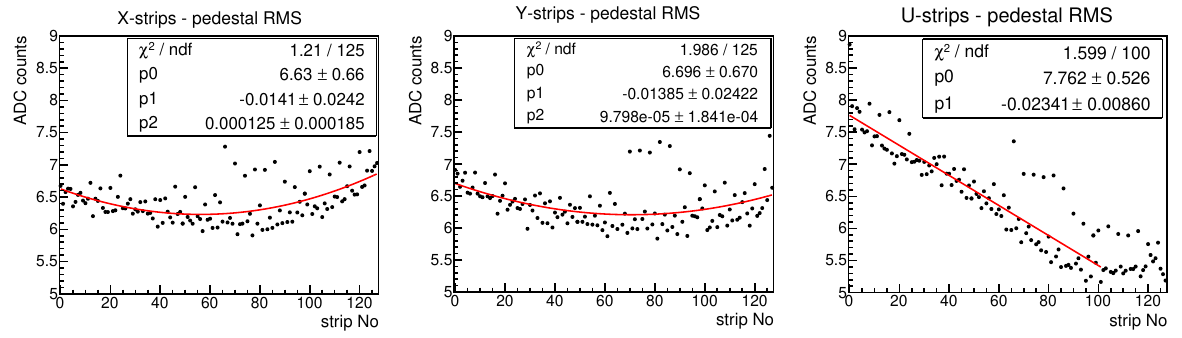}
\caption{\label{fig:pedestals} Plots of the pedestal noise (rms ADC counts) of the channels of APV25 FE cards connected to the three strips. The solid red line is a polynomial function of the order 2 fit to the X-strips and Y-strips data points.} 
\end{figure}

The plots of the rms pedestal fluctuation distributions of the X, Y and U strips of the XYU-GEM prototype readout planes are presented in Fig.~\ref{fig:pedestals}. The rms value of the distributions represents the pedestal noise in ADC counts for each APV25 channel. The first two plots on the left and center show the rms values for each of the 128 channels in X-strips and Y-strips respectively.  In this case, where all the strips have the same length of 100 mm,  the rms values fluctuate around 6.5 ADC counts for all channels. The slightly curved profile of the values with respect to the APV25 channel number reflects the routing of the 128 channels of the APV25 chip to the 130-pin connectors on the FE card. The red line is a polynomial function of the order 2 fitted to the experimental data.
The rms values for the U-strips in the right plot in Fig.~\ref{fig:pedestals} show a clear linear dependence with the strip number, varying from $\sim$7.5 ADC counts for the longest strips at the diagonal axis to $\sim$5.5 ADC counts for channels connected to the shortest strips ($\ge$ strip number 100) when moving toward the corner of the readout plane. A straight line fit through the data indicates the correlation between strip length and hence the input capacitance of the U-strips as seen by the APV25 channel and the rms noise of the given channel. The  remaining 28 strips are too short and  provide a negligible contribution to the overall pedestal rms.
%
%
\subsection{Efficiency vs. HV}
\label{subsec:eff}
\begin{figure}[!ht]
\centering
\includegraphics[width=1\columnwidth,trim={0pt 0mm 0pt 0mm},clip]{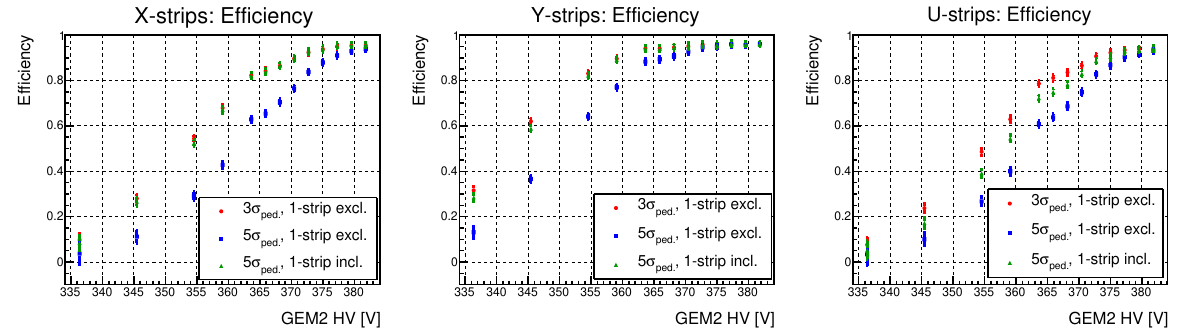}
\caption{\label{fig:xyugem_eff_hv} Plots of the efficiency for X-strips \textit{(left)}, Y-strips \textit{(center)} and U-strips \textit{(right)} as a function of the voltage on GEM2; red, blue and green markers show data for \textcolor{red}{3$\rm \sigmaup_{ped}$, 1-strip excl.},  \textcolor{blue}{5$\rm \sigmaup_{ped}$, 1-strip excl.} and \textcolor{ForestGreen}{5$\rm \sigmaup_{ped}$, 1-strip incl.} respectively.}
\end{figure}

The plots of Fig.~\ref{fig:xyugem_eff_hv} are the efficiency curves for X-strips, Y-strips and U-strips  as a function of the average voltage (HV) applied across the three GEM foils. This also corresponds to the voltage across the second GEM foil (GEM2) when the COMPASS resistive divider~\cite{COMPASS:2002} is used. For the results presented in the current study, we examined the resulting efficiency achieved based on three different conditions which combined pedestal cuts and a minimum number of strips per cluster requirement. The conditions tested are:
\begin{itemize}
\item 3$\rm \sigmaup_{ped}$ pedestal cut and a minimum two strips per cluster requirement i.e. single-strip clusters events excluded from the analysis - \textcolor{red}{(3$\rm \sigmaup_{ped}$, 1-strip excl.)}.
\item 5$\rm \sigmaup_{ped}$ pedestal cut and a minimum two strips per cluster requirement i.e. single strip clusters events excluded  from the analysis - \textcolor{blue}{(5$\rm \sigmaup_{ped}$, 1-strip excl.)}.
\item 5$\rm \sigmaup_{ped}$ pedestal cut and single-strip cluster events included in the analysis - \textcolor{ForestGreen}{(5$\rm \sigmaup_{ped}$, 1-strip incl.)}.
\end{itemize}

The efficiency in each plane is defined as the ratio between the number of events with at least one cluster in given plane of the XYU-GEM prototype and the total number of events in which a track is reconstructed from hits recorded simultaneously in both the x-axis and y-axis for all four GEM trackers. Additionally, only the events in the XYU-GEM prototype with the hit coordinates within a 500 $\muup$m radius of the projected coordinates from the fitted tracks are considered in the efficiency analysis.
For all three conditions the efficiency of the Y-strips on the GEM2 plateaus at 96\% for voltages higher than 372 V (middle plot of Fig.~\ref{fig:xyugem_eff_hv}) for all three conditions of pedestal cuts and minimum number of strips.  The plateau is reached at higher voltage, $\sim$380 V for X-strips and U-strips because of the lower signal on these strips which we will discuss in the following section~\ref{subsec:adc}. The efficiency drop at lower HV settings is more severe for the condition 5$\rm \sigmaup_{ped}$ cuts and single-strip clusters excluded \textit{(blue squares)} in  all three planes because, at lower voltages, when the 5$\rm \sigmaup_{ped}$ cut is applied a significant number of events will be single-strip cluster events. It is  noteworthy  that the difference in efficiency for the other two conditions i.e. the 3$\rm \sigmaup_{ped}$ cut with single-strip cluster events excluded  \textit{(red dots)} and the 5$\rm \sigmaup_{ped}$ cut with single-strip cluster events included \textit{(green triangles)}, is negligible over all voltage settings for all three X-strips, Y-strips, and U-strips. This important result means one can apply a 3$\rm \sigmaup_{ped}$ pedestal cut to achieve high a level of efficiency while minimizing the probability of false single-strip events by requiring a minimum two-strip cluster events.
%
%
%
\subsection{Cluster charge vs. HV}
\label{subsec:adc}
\begin{figure}[!ht]
\centering
\includegraphics[width=1. \columnwidth,trim={0pt 0mm 0pt 0mm},clip]{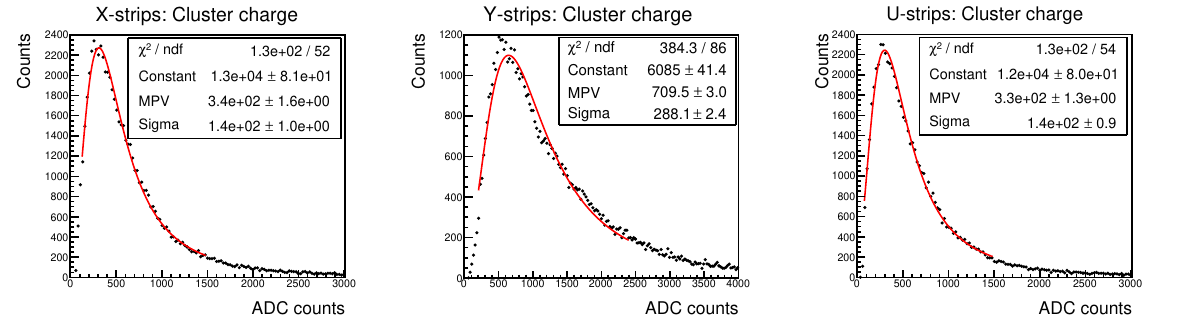}
\caption{\label{fig:xyugem_adc_dist} Cluster charge distribution for Y-strips \textit{(left)}, U-strips \textit{(center)} and X-strips \textit{(right)} for an average voltage of 382 V on the GEMs. The data are fitted to a Landau function (red).}
\end{figure}

 Plots of the cluster charge distributions in ADC counts for X-strips and Y-strips and U-strips for HV = 382 V on GEM2 are shown in  Fig.~\ref{fig:xyugem_adc_dist}.  The distribution for each readout plane is fitted to a Landau function as shown in red in the plots and the  most probable value (MPV) of the Landau fit is used as the average cluster charge at each HV setting.  The cluster charge is defined as the sum of the charge of all strips in a cluster, i.e consecutive strips above a chosen pedestal cut (n$\rm \sigmaup_{ped}$) of the given channel. The strip charge is the  integrated charge of all nine APV25 time samples.
As shown in the plots of Fig.~\ref{fig:xyugem_adc_dist}, the average cluster charge for Y-strips is 709.5 $\pm$ 3 ADC counts which is more than twice the average cluster charge for for X-strips (341.3 $\pm$ 1.6 ADC counts) and U-strips (329.3 $\pm$ 1.3 ADC counts).  This significant difference in charge collection between the Y-strips on the one hand and for X-strips and U-strips on the other hand is explained by a sub-optimal design of the (X--Y--U)-strip readout. This is because the capacitance coupling of the Y-strips on layer L3 with the pad of layer L2 is stronger than X-strips and U-strips on layer L4 as the gap between layers L2 and L3 is half of the gap between layers L2 and L4 (See Fig~\ref{fig:capaShStripxyu}). Possible ways to achieve a more equal collection of the charges between strips on L3 and L4 are to reduce width of the strips on L3 or increase the gap between layers L2 and L3. 
We plan to perform simulation studies to optimize strip width and dielectric thickness for the X-strips, Y-strips and U-strips using tools such as Garfield++~\cite{garfield} and COMSOL Multiphysics~\cite{comsol} to achieve equal sharing of the charges between the three set of strips. The optimized parameters will be implemented in future (X--Y--U)-strip readout prototypes.
\begin{figure}[!ht]
\centering
\includegraphics[width=1\columnwidth,trim={0pt 0mm 0pt 0mm},clip]{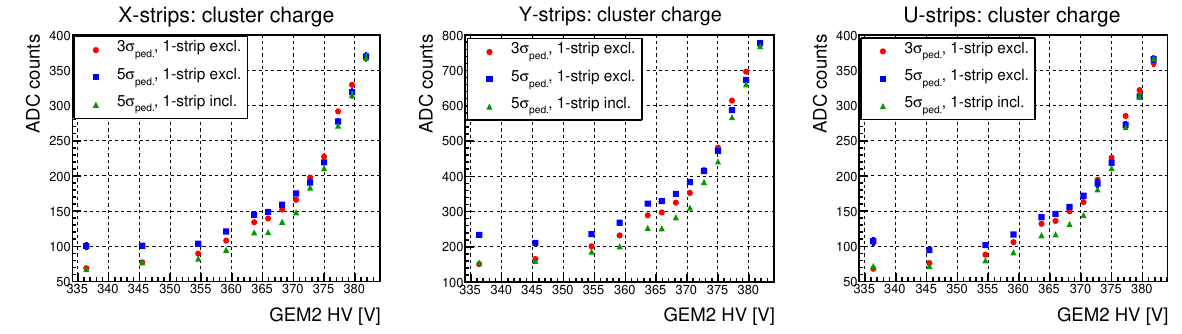}
\caption{\label{fig:xyugem_adc_hv} Plots of average cluster charge  (in ADC counts) for X-strips \textit{(left)}, Y-strips \textit{(center)} and U-strips \textit{(right)}  as a function of the voltage on GEM2; red, blue and green markers show data for \textcolor{red}{3$\rm \sigmaup_{ped}$, 1-strip excl.},  \textcolor{blue}{5$\rm \sigmaup_{ped}$, 1-strip excl.} and \textcolor{ForestGreen}{5$\rm \sigmaup_{ped}$, 1-strip incl.} respectively.}
\end{figure}
The plots of Fig.~\ref{fig:xyugem_adc_hv} show the average cluster charge as a function of the voltage on GEM2 for X-strips (left), Y-strips (center) and U-strips (right).  As already mentioned, Y-strips collect  $\sim$50\% of the total cluster charge while X-strips and U-strips  collect $\sim$25\% of the total cluster charge each for all HV settings. The average cluster charge is higher at the lower HV values when 5$\rm \sigmaup_{ped}$ cut and 1-strip cluster events excluded (blue squares) are applied.
%
\subsection{Cluster charge correlation}
\label{subsec:chargeSh}
\begin{figure}[!ht]
\centering
\includegraphics[width=1\columnwidth,trim={0pt 0mm 0pt 0mm},clip]{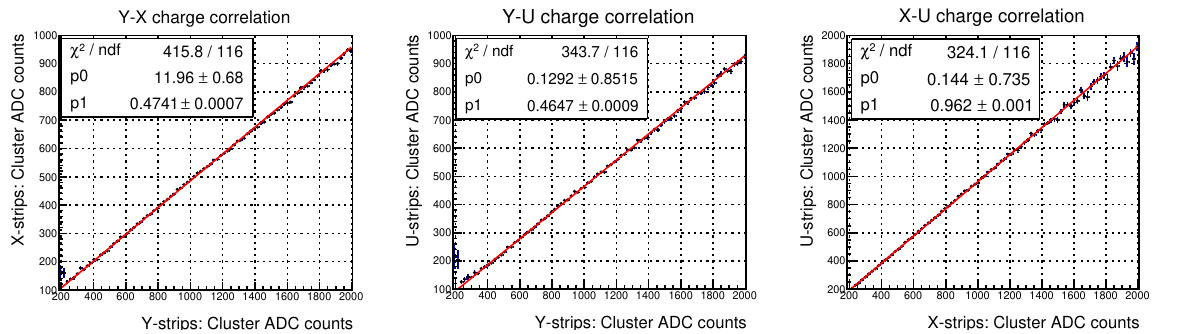}
\caption{\label{fig:xyugem_chargeSh} Charge sharing correlation between different sets of two readout planes for a HV = 382 V on GEM2; \textit{(left)} x-plane vs. y-plane;  \textit{(center)} u-plane vs. y-plane;  \textit{(right)} x-plane vs. u-plane.}
\end{figure}

The plots of Fig.~\ref{fig:xyugem_chargeSh} illustrate the excellent cluster charge correlation between the Y-strips and X-Strips \textit{(left)}, the Y-strips and U-strips \textit{(center)} and the X-strips and U-strips \textit{(right)}. The data points in each of the correlation plots are fitted to straight line functions with the slopes representing the charge ratio between the strips of two given planes. The charge  ratios for Y:X = 0.474, X:U = 0.962 and Y:U = 0.464 are consistent with the observation made on the results of section~\ref{subsec:eff} and~\ref{subsec:adc}. 
%
%
\subsection{Strip multiplicity vs. HV}
\label{subsec:stripMult}
\begin{figure}[!ht]
\centering
\includegraphics[width=1.0 \columnwidth,trim={0pt 0mm 0pt 0mm},clip]{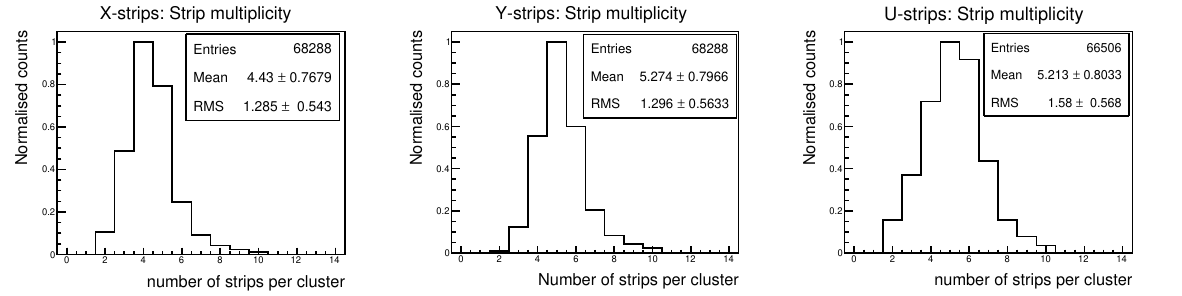}
\caption{\label{fig:xyugem_stripMult_dist} Histograms of trip multiplicity for 3$\rm \sigmaup_{ped}$ cut and a minimum of 2 strips per strip cluster for X, Y and U strips.}
\end{figure}

The strip multiplicity of an event is defined as the number of neighboring strips (i.e. a strip cluster) with ADC counts (charge) above the chosen pedestal cut. Fig.~\ref{fig:xyugem_stripMult_dist} shows the distribution of strip multiplicity for X-strips  \textit{(left)} and in Y-strips  \textit{(center)} and U-strips  \textit{(right)} planes for an HV = 382 V on GEM2, with a 3$\rm \sigmaup_{ped}$  cut and minimum 2-strip cluster requirement (i.e. single-strip cluster events excluded). The average strip multiplicity is 4.43 $\pm$ 0.77,  5.27 $\pm$ 0.8,  and 5.21 $\pm$ 0.8 for X-strips, Y-strips and U-strips respectively. The  difference of $\sim$0.84  between X-strips and Y-strips is due to the larger signal on Y-strips compared to X-strips, whereas the larger strip multiplicity for U-strips, 5.21 compared to  X-strips is because along the diagonal axis, the actual pitch of U-strips plane is equal to 567 $\muup$m  (800 $\muup$m / $\sqrt{2}$) as opposed to 800  $\muup$m for X-strips. 
\begin{figure}[!ht]
\centering
\includegraphics[width=1.0\columnwidth,trim={0pt 0mm 0pt 0mm},clip]{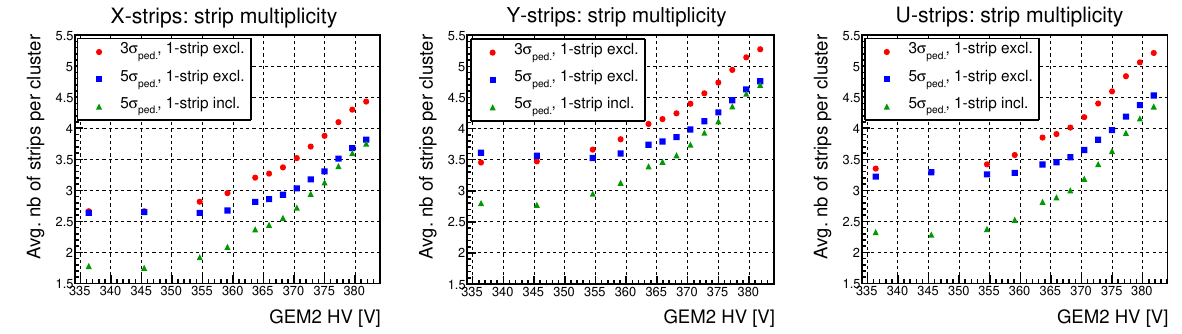}
\caption{\label{fig:xyugem_stripMult_scan} Plots of the average strip multiplicity for X, Y and U strips as a function of the applied voltage on GEM2; red, blue and green markers indicate data points for \textcolor{red}{3$\rm \sigmaup_{ped}$, 1-strip excl.},  \textcolor{blue}{5$\rm \sigmaup_{ped}$, 1-strip excl.} and \textcolor{ForestGreen}{5$\rm \sigmaup_{ped}$, 1-strip incl.} respectively.}
\end{figure}
The plots in  Fig.~\ref{fig:xyugem_stripMult_scan} are the average strip multiplicity as a function of the voltage on GEM2 for X-strips  \textit{(left)} and in Y-strips  \textit{(center)} and U-strips  \textit{(right)} planes and for the three cut settings described in section~\ref{subsec:eff}. The average strip multiplicity of the X-strips, Y-strips and U-strips ($\sim$2.6, $\sim$3.5 and $\sim$3.3 respectively) is almost independent of the applied voltage for voltages $\leq$ 355 V with requirement of two strips per cluster and the applied pedestal cuts of 3$\rm \sigmaup_{ped}$ and 5$\rm \sigmaup_{ped}$ for all three readout planes. For  voltages $\ge$ 355 V, the average strip multiplicity increases linearly with the applied voltage and as expected is slightly higher for the 3$\rm \sigmaup_{ped}$ cut than the 5$\rm \sigmaup_{ped}$ cut. When single-strip cluster events are included in the analysis, the average strip multiplicity is significantly smaller at lower HV values because of the higher probability of single-strip cluster events. At the higher HV values, the average strip multiplicity for 5$\rm \sigmaup_{ped}$ cuts, converge for both single-strip cluster events and for events when a minimum 2-strip is required in the cluster, because the probability of single-strip events becomes negligible for the X-strips and Y-strips and small for U-strips.

 \section{Spatial resolution studies}
\label{sec:resolution}
For the spatial resolution studies, clean tracks were selected by requiring correlated hits from both the X and Y planes in  all four GEM trackers simultaneously. The tracks are fitted to a straight line function, independently for X and Y planes. Hits from the XYU-GEM prototype under study are excluded from the track fit reconstruction in order to avoid any ambiguity or bias in the track residuals analysis. The track residual for X-strips, Y-strips and U-strips planes is the difference between the projected coordinates $\rm x_{Fit}$, $\rm y_{Fit}$ and $\rm u_{Fit}$ of the fitted tracks at the z position of the XYU-GEM prototype plane and at the measured hit coordinates $\rm x_{Meas}$, $\rm y_{Meas}$ and $\rm u_{Meas}$ from test beam data within a 500 $\muup$m radius search area around the projected coordinates of the fitted tracks. The  coordinates $\rm u_{Fit}$ are calculated directly from both x$_{Fit}$ and $\rm y_{Fit}$ coordinates of the fitted tracks.
\begin{figure}[!ht]
\centering
\includegraphics[width=1 \columnwidth,trim={0pt 0mm 0pt 0mm},clip]{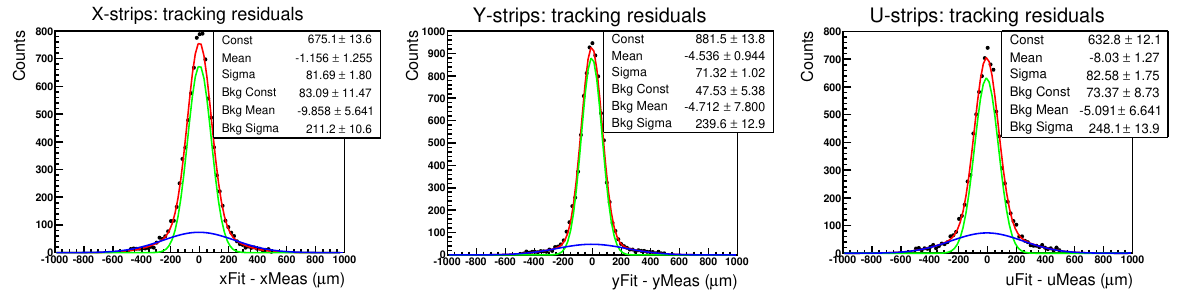}
\caption{\label{fig:xyugem_residuals} Plots of tracking residual distribution for X, Y and U strips for HV = 382V on GEM2. The data (black dots) are fitted to a double Gaussian function  (red curves). The green curves are the Gaussian function with narrow width and the blue curves the Gaussian function with wider width.}
\end{figure}
Fig.~\ref{fig:xyugem_residuals} shows the plots for the track residual distributions  for X-strips, Y-strips and U-strips. A double Gaussian function is fitted to the residual data and the widths of the narrower Gaussian are $\rm \sigmaup_{x}^{res}$ = 78.5 $\pm$ 1.2 $\muup$m for X-strips,  $\rm \sigmaup_{y}^{res}$ = 70.37 $\pm$ 0.7 $\muup$m for Y-strips and  $\rm \sigmaup_{u}^{res}$ = 81.29 $\pm$ 1.3 $\muup$m for U-strips. As expected, the width of the residual distribution is about 10 $\muup$m smaller for the Y-strips than for the X-strips and U-strips because of the larger signal-to-noise ratio.
\begin{figure}[!ht]
\centering
\includegraphics[width=1\columnwidth,trim={0pt 0mm 0pt 0mm},clip]{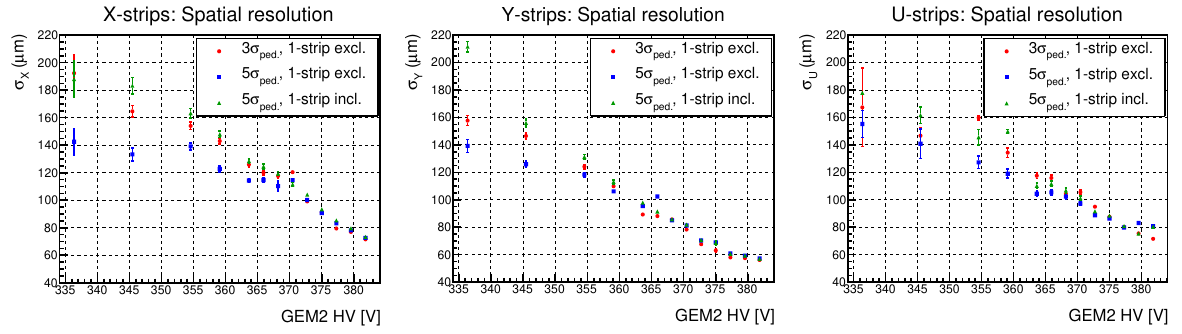}
\caption{\label{fig:xyugem_resol_hv_scan}   Plots of the spatial resolution  for X, Y and U strips \textit{(right)}  as a function of the voltage on GEM2; red, blue and green markers indicate the data points for \textcolor{red}{3$\rm \sigmaup_{ped}$, 1-strip excl.},  \textcolor{blue}{5$\rm \sigmaup_{ped}$, 1-strip excl.} and \textcolor{ForestGreen}{5$\rm \sigmaup_{ped}$, 1-strip incl.} respectively.}
\end{figure}
The tracking telescope in this study is identical to the one presented in a previous work~\cite{capaSh_urwell2022} and the same approach described in the paper is used to calculate the values of the track fit error $\rm \sigma_{x}^{err}$ and $\rm \sigma_{y}^{err}$ associated to the fitted tracks for X-strips and Y-strips respectively at the z location of (X--Y--U)-GEM prototype. The computed errors are $\rm \sigma_{x}^{err}$ = (31.2 $\pm$ 0.6) $\muup$m for X-strips and $\rm \sigma_{y}^{err}$ = (30.6 $\pm$ 0.6) $\muup$m for Y-strips. The error $\rm \sigma_{u}^{err}$ for U-strips is the average $\rm 0.5 \times \, (\sigma_{x}^{err} \, +  \sigma_{y}^{err})$ = (30.9 $\pm$ 0.6) $\muup$m of the error in x and in y.  The spatial resolution in each plane is then obtained by subtracting in quadrature the track fit error values from the width of the Gaussian fit to the residual distribution.  Fig.~\ref{fig:xyugem_resol_hv_scan} presents plots of the spatial resolution of X, Y and U strips as a function of voltage on GEM2 for the three cut settings described in subsection~\ref{subsec:eff}. The best performance is achieved for HV = 382 V on GEM2 and for 3$\rm \sigmaup_{ped}$ cut and minimum 2-strip cluster requirement, resulting in a spatial resolution of $\rm \sigmaup_{x}^{res}$ = 71.6 $\pm$ 0.8 $\muup$m for X-strips,  $\rm \sigmaup_{y}^{res}$ = 56.2 $\pm$ 0.9 $\muup$m for Y-strips and $\rm \sigmaup_{u}^{res}$ = 75.2 $\pm$ 0.9 $\muup$m for U-strips. For higher voltages (HV $ \rm \ge 363 V$),  the spatial resolution no longer depends on the pedestal cuts and minimum number of strips in a cluster, except for U-strips at the highest HV settings when we observe a small worsening of the resolution for the    5$\rm \sigmaup_{ped}$ data. The reason for the degradation is not obvious to us but we suspect some issues with the raw APV25 data  during data acquisition. At lower voltages, when the detector gain is low, a large difference in spatial resolution performance is observed for 5$\rm \sigmaup_{ped}$ data between the single-strip and 2-strip cluster requirement and for 2-strip clusters data between 3 and 5$\rm \sigmaup_{ped}$ cuts because of the more significant impact of these requirements on signal-to-noise ratio for each strip plane. The optimization of the (X--Y--U) strips readout planes for future prototypes as discussed in subsection~\ref{subsec:chargeSh}, will ensure a more equal charge sharing  between the three X-strips,Y-strips and U-strips planes and a more uniform spatial resolution performance of better than 60 $\muup$m.

\section{Summary and future work}
\label{sec:conclusion}
The performance of a 10 cm $\times$ 10 cm triple-GEM prototype  with 3-coordinates (X--Y--U)-strip capacitive-sharing anode readout was evaluated in test beam studies at Jefferson Lab. The new readout structure allows simultaneous measurement of three coordinates of particle interaction in a single detector and provides a powerful tool to address multi-hit ambiguity and pattern recognition challenges. These challenges occur when large MPGD tracking detectors are operated in low to moderate particle rate environments such as expected for the EIC central detector or  the tracking systems of the large spectrometer installed in the Jefferson Lab experimental Halls B and D. 
\begin{figure}[!ht]
\centering
\includegraphics[width=0.65\columnwidth,trim={0pt 0mm 0pt 0mm},clip]{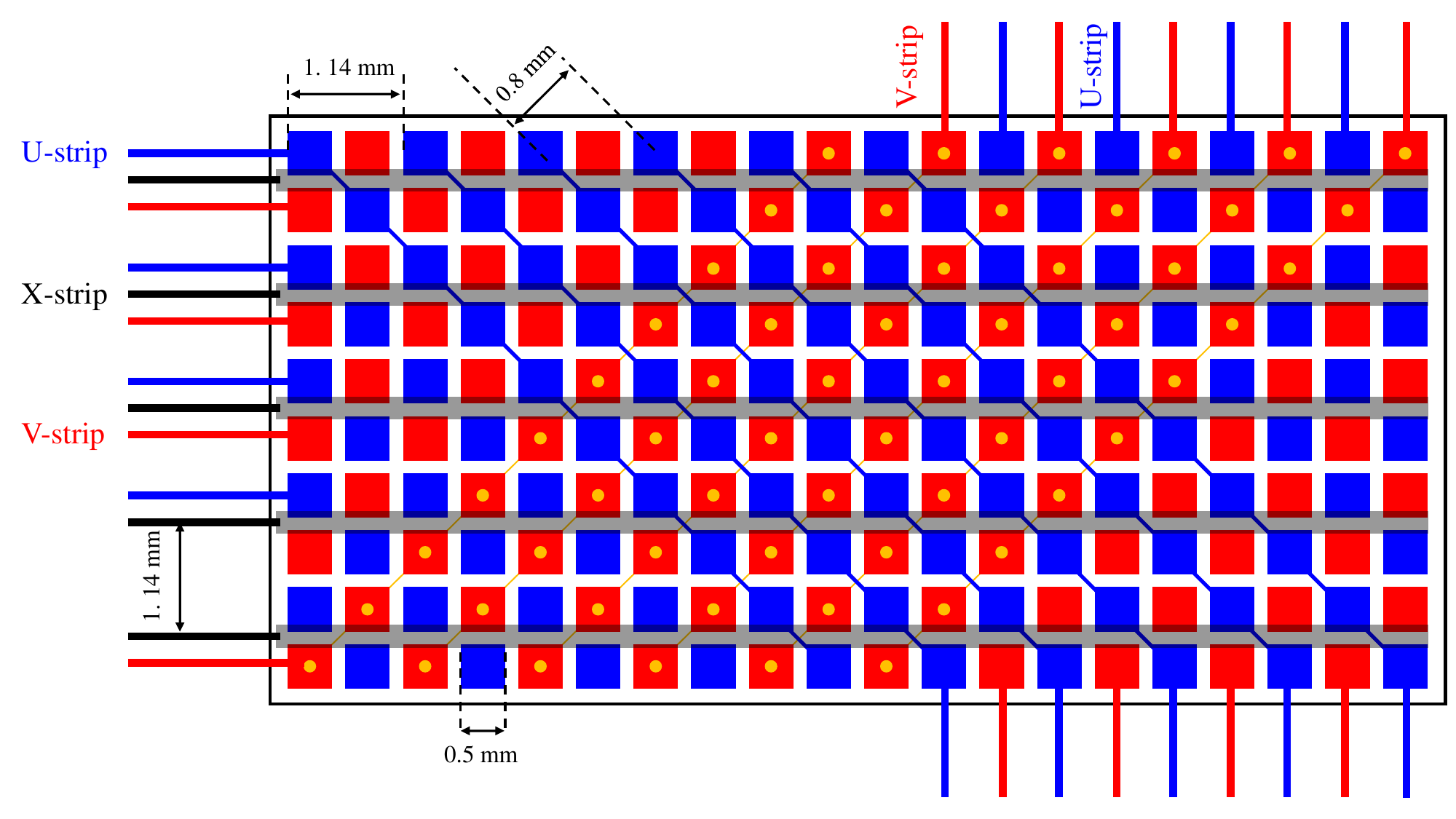}
\caption{\label{fig:uxv_strip} Conceptual design of 3-coordinates U-X-V-strip layer of capacitive-sharing readout structures for large area $\muup$RWELL.}
\end{figure}
The prototype was fabricated and successfully tested in the electron beam of the Pair Spectrometer facility in experimental Hall D at Jefferson Lab. We demonstrated that a spatial resolution better than 60 $\muup$m could be achieved for Y-strips plane and  $\sim$75 $\muup$m for X-strips and U-strips planes at full efficiency with a strip pitch of 800 $\muup$m. Design modifications of the (X--Y--U)-strip readout PCB have been identified to improve the performance and response uniformity of all three strip planes. The strip widths of the three strip planes as well as the gap between the capacitive and strip layers of the capacitive-sharing structure need to be optimized to ensure equal charge sharing among the strip planes.  An additional optimization of the pad size of the capacitive-sharing layer is also necessary to limit the average strip hit multiplicity to 3 strips per clusters instead of 5 strips per cluster that was obtained with the current prototype.  A large pitch strip readout optimized for a 3-strip cluster could achieve similar spatial resolution while reducing the signal size by 40\% and improving the rate capability of the detector accordingly. Another critical challenge of capacitive-sharing strip readout for large area MPGD trackers is to minimize the detector input capacitance seen by front end electronics connected to the readout strips. This is needed in order to maintain good signal quality and large signal-to-noise ratio. 

Dedicated R\&D efforts are ongoing to develop low-capacitance long and large strip readout PCB for MPGDs as well as to explore design variations of the 3-coordinate capacitive-sharing strip readout concept. One such variation is the (U-X-V)-strip design shown in Fig.~\ref{fig:uxv_strip} under development as part of the R\&D program of large $\muup$RWELL tracker for future EIC second detector. These modifications will be implemented in future prototypes. However, large strip pitch readout structures such as the one discussed in this current study are not well suited for high rate environments such as expected for the high luminosity LHC detectors. A signal spread over 5 $\times$ 800 $\muup$m-pitch strips in a high rate application will result in a cluster size of 4 mm which makes it challenging in avoiding electronic pile-up and resolving multiple-track situations. In this case, smaller pitch readouts and fast integration electronics are still necessary to achieve adequate spatial resolution.

\section{Acknowledgements}
\label{sect_acknowledgements}
This material is based upon work supported by the U.S. Department of Energy, Office of Science, Office of Nuclear Physics under contracts DE-AC05-06OR23177.\\
The prototype was tested in beam in Hall D  at Jefferson Lab. We thank the lab and Hall D managements (in particular L. Pentchev and E. Chudakov) for their continuing support and the technical team for the tremendous help in integrating our test setup in the Hall D test beam area. Finally I would like to thank Ms. Nicole Wingnannon for the invaluable support.

\newpage
\bibliographystyle{elsarticle-num}
\nocite{*}
\bibliography{sec6_reference.bib}
\end{document}